\documentclass[
 reprint,
 amsmath,amssymb,
 aps,
pra,
floatfix]{revtex4-2}

\usepackage{graphicx}
\usepackage{dcolumn}
\usepackage{bm}
\usepackage{amsmath}
\usepackage{physics}

\newcommand{\up}{\uparrow}
\newcommand{\dow}{\downarrow}

\newcommand{\1}{\uparrow} 
\newcommand{\2}{\downarrow} 
\newcommand{\be}{\begin{equation}}
\newcommand{\ee}{\end{equation}}

\begin{document}

\preprint{APS/123-QED}

\title{Saturating interaction in coherently coupled two-component Bose-Einstein condensates}
%
\author{R. Eid}
\author{S. Tiengo}
\author{M. L\'evy}
\author{T. Bourdel}
\thanks{thomas.bourdel@institutoptique.fr}
\affiliation{%
 Universit\'e Paris-Saclay, Institut d'Optique Graduate School,\\
 CNRS, Laboratoire Charles Fabry, 91127 Palaiseau, France 
}%

\date{\today}

\begin{abstract}
Rabi-coupled spinor Bose-Einstein condensates, with competing intra- and interspecies interactions, enable independent control of two- and three-body interactions. We show that coupling can also drive the system into a strongly nonlinear regime of saturating interaction. More precisely, the equation of state interpolates between low- and high-density regimes described by two different two-body scattering lengths. Interestingly, the transition can be determined by the strength of the coupling. We experimentally demonstrate this saturation phenomenon by measurements of the interaction energy of a Bose-Einstein condensate as a function of the detuning and of the strength of the Rabi coupling in spin mixtures of potassium 39. 
\end{abstract}

\maketitle
\section{Introduction}
For small perturbations, a physical system usually responds according to linear response theory. As the perturbation strength increases, the response may become non-linear giving rise to a wide variety of physical phenomena (solitons, bistability, chaotic behavior, etc...). \cite{Strogatz2024}. Nonlinear response is often modeled by the addition of other response terms in powers of the perturbation. Such a Taylor expansion approach explains the higher harmonic generation in a system driven at a single frequency, as is commonly observed in nonlinear optics \cite{Shen1984}. The Taylor expansion of the nonlinear response is not always valid for large perturbations, as many physical systems exhibit saturation of their response at large driving amplitudes. Typical examples are the magnetic or electric responses of materials that saturate when the maximum polarization is reached. In optics, saturation of the non-linear index of refraction, of absorption, or of the gain in amplifiers is also commonly observed \cite{Boyd2020}. 

In the physics of dilute atomic Bose-Einstein condensates, the non-linearity primary comes from the interaction between atoms \cite{Pitaevskii2003} which are characterized by the scattering length $a$. In the case of a single-component Bose-Einstein condensate in the weak interaction regime $na^3 \ll 1$, where $n$ is the atomic density, the energy density is dominated by the mean-field interaction scaling as $n^2$. Beyond this paradigm, quantum fluctuations are responsible for the Lee-Huang Yang term scaling non-analytically as $n^{5/2}$ in the energy density \cite{Lee1957}. In bosonic mixtures, this term is responsible for the stabilization of quantum droplets \cite{petrov2015, cabrera2018}. Higher-order interaction terms, for example, corresponding to a three-body interaction, are usually negligible in the dilute regime $na^3 \ll 1$ \cite{Braaten1997, Kohler02, Mestrom19}.

However, this paradigm is no longer valid in the presence of coupling between two spin states, i.e. when the condensate is in a dressed state that is a coherent superposition of two spin states. A condensate in a dressed state may experience both reduced usual two-body interactions \cite{sanz2022} and large three-body interactions that are due to possible virtual excitations during collisions \cite{Tiengo2025}. Such three-body processes are increasingly relevant as the ratio $\gamma$ of the two-body interaction strength to the Rabi coupling strength increases.  Their effect on the equation of state, that is, the addition of a $n^3$ term, was experimentally measured through modifications in the condensate dynamics \cite{Hammond2022}. 

In this paper, we consider the situation where $\gamma \gtrsim 1$ and demonstrate a saturation of the interaction energy as a function of $\gamma$, an effect that can be captured in a mean field framework. In this regime of strong nonlinear behavior, the equation of state can no longer be described as a Taylor series expansion as a function of the density, with terms scaling as $n^p$ describing $p$-body interactions. In contrast, one needs to compute the exact solution.  

The paper is organized as follows. We first describe the coupled system Hamiltonian and provide a theoretical explanation for the saturation of the interaction energy in a homogeneous system (with a fixed density $n$). Second, we present our experiment using $^{39}$K atoms. We explain the preparation of the condensates in the dressed state and the measurement of the interaction energy through one-dimensional time-of-flight expansions. As the condensate lies in the one-dimensional (1D) to three-dimensional (3D) crossover regime, we finally compare our results with Gross-Pitaevskii simulations that include the two spin states, the Rabi coupling, as well as the radial trap. A good match, with no fit parameter, confirms our interpretation of the data. 

\section{Theoretical description}

We consider a Bose gas in a volume $V$ consisting of $N$ atoms of mass $m$ with two coupled internal states, $\sigma = \1,\2$. 
The three relevant interaction constants $g_{\sigma\sigma'}=4\pi\hbar^2a_{\sigma\sigma'}/m$ are related to the scattering lengths $a_{\sigma\sigma'}$ and $\hbar$ the reduced Planck constant. In an homogeneous system with density $n=N/V$, the mean field energy for a condensate in the spinor state $(\phi_\1, \phi_\2)$ reads
\begin{gather}
\begin{split}
\frac{E_{MF}}{V}=-\frac{\hbar\Omega}{2}(\phi_\1^*\phi_\2+\phi_\2^*\phi_\1)+\frac{\hbar\delta}{2}(\left|\phi_\1\right|^2-\left|\phi_\2\right|^2)\\
+\sum_{\sigma\sigma'}\frac{g_{\sigma\sigma'}}{2}\left|\phi_\sigma\right|^2 \left|\phi_{\sigma'}\right|^2 \textrm{,}
\end{split}
\end{gather}
where $\delta$ is the coupling detuning and $\Omega$ the Rabi coupling strength. 
The ground state is found upon minimization of the
energy with respect to the internal state. The first term fixes the relative phase of the spinor that we can thus write $(\phi_\1, \phi_\2)=\sqrt{n}(\sin(\theta/2), \cos(\theta/2))$ with $\theta \in [0, \pi]$. The energy is then
\begin{gather}
\frac{E_{MF}}{N}=-\frac{\hbar\Omega}{2} \sin\theta-\frac{\hbar\delta}{2}\cos \theta\nonumber\\+\frac{gn}{2}-\frac{\Delta g n}{4}\cos{\theta}-\frac{\overline{g}n}{2}\sin^2\theta, \label{eqEMF}
\end{gather}
where $g=(g_{\1\1}+g_{\2\2})/2$, $\Delta g=g_{\1\1}-g_{\2\2}$, and $\bar{g}=(g_{\1\1}+g_{\2\2}-2g_{\1\2})/4$.

The minimization of Eq.\,\ref{eqEMF} as a function of $\theta$ leads to the following implicit equation for the spin composition
\begin{gather}
\cot\theta=\frac{\delta}{\Omega}+\alpha -2\gamma \cos\theta \textrm{,}\label{Eq:spin}
\end{gather}
where $\alpha=\frac{\Delta g n}{2\hbar\Omega}$ and $\gamma=\frac{\bar{g}n}{\hbar \Omega}$. The spin magnetization and, equivalently, the mixing angle $\theta$ thus depend not only on $\delta/\Omega$ but also on the density $n$. In the low-density regime, i.e. for $\alpha, \gamma\ll1$, the density dependence of the spin composition can be calculated in perturbation. This effect was experimentally observed \cite{Farolfi2021}. At the same order in perturbation, we find two- and three-body corrections on the interaction energy, generalizing the results from \cite{Hammond2022} to the asymmetric case
\begin{gather}
\frac{E_{MF}}{N}\approx \epsilon_-+g_2\frac{n}{2}+g_3\frac{n^2}{3} \label{energy}\\ \nonumber
\text{with } \epsilon_-=-\hbar\sqrt{\Omega^2+\delta^2}/2,\\ \nonumber
g_2=g-\frac{\overline{g}}{1+\delta^2/\Omega^2}-\frac{\Delta g}{2} \frac{\delta/\Omega}{\sqrt{1+\delta^2/\Omega^2}},\\ \nonumber
g_3=-\frac{3\overline{g}^2}{\hbar\Omega}\frac{\delta^2/\Omega^2}{\left(1+\delta^2/\Omega^2\right)^{5/2}}+\frac{3\bar{g}\Delta g\delta/\Omega}{2 \hbar \Omega \left(1+\delta^2/\Omega^2\right)^{2}}\\ \nonumber -\frac{3\Delta g^2}{16 \hbar \Omega\left(1+\delta^2/\Omega^2\right)^{3/2}}.
\end{gather}
$\epsilon_-$ is the one-body Rabi energy of the lower-dressed state $\ket{-}$. 
The two-body coupling constant $g_{2}(\delta/\Omega)$ corresponds to the interaction between two atoms in $\ket{-}$. It interpolates between $g_{\1\1}$ and $g_{\2\2}$, reaching intermediately a minimum $g_{\rm{min}}=(g_{\1\1}g_{\2\2}-g_{\1\2}^2)/4\bar{g}$ in the experimentally relevant case $g_{\up \dow}<g_{\up\up}, g_{\dow \dow}$. 
The variation of this two-body term was experimentally measured in \cite{sanz2022}. The three-body coupling constant $g_3(\delta/\Omega, \Omega)$ scales as $1/\Omega$ and was also experimentally demonstrated \cite{Hammond2022}.

\begin{figure}
\includegraphics[width=\columnwidth]{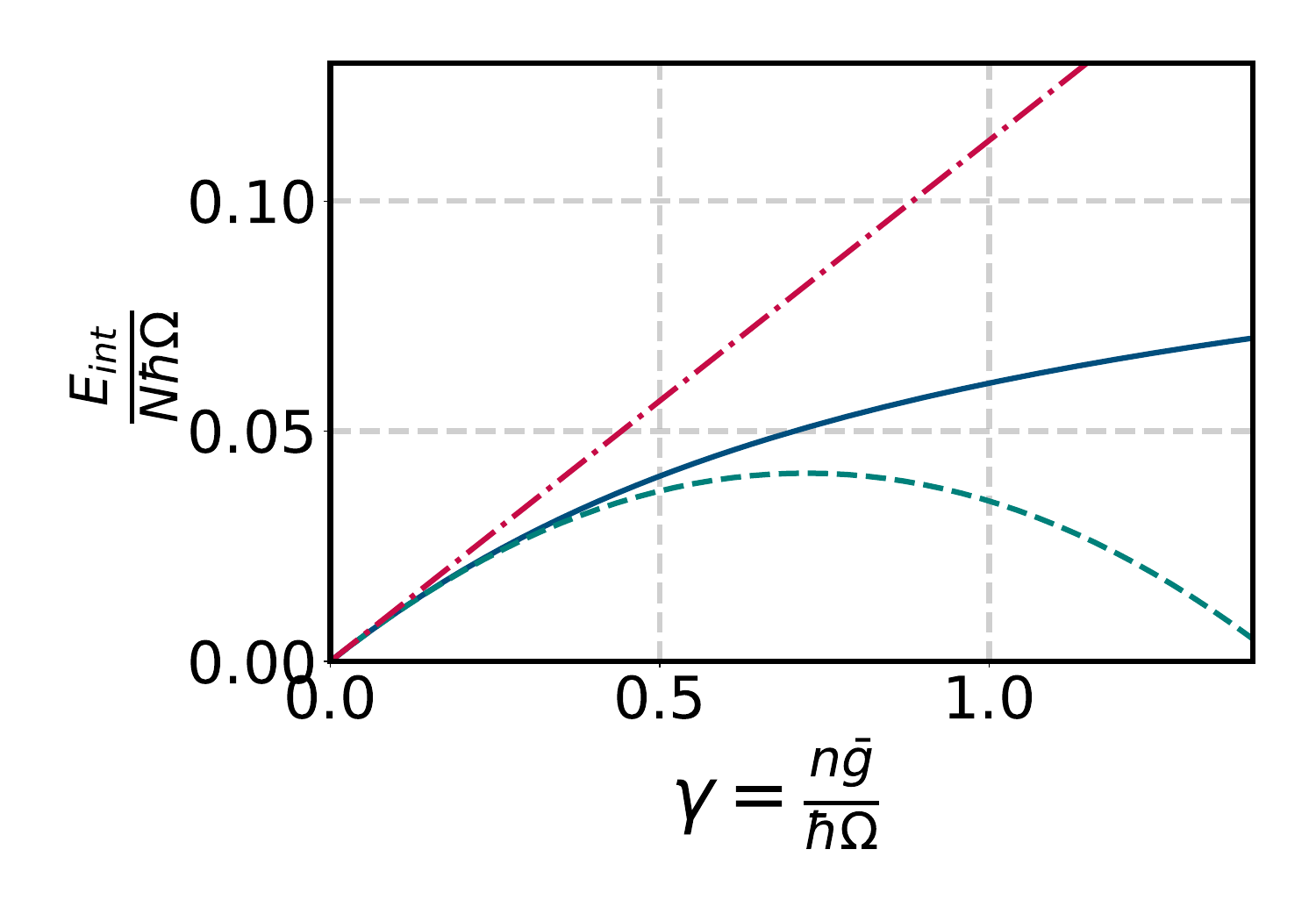}
\caption{\label{fig:interaction} 
Interaction energy per particle $E_{\rm{int}}/N=(E_{\rm{MF}}/N-\epsilon_-)$ as a function of $\gamma$. The specific parameters are $\delta/\Omega=1$, $a_{\up \up}=86.8\,a_0$, $a_{\dow \dow}=33.3\,a_0$, and $a_{\up \dow}=-53.2\,a_0$, where $a_0$ is the atomic Bohr radius. The solid blue curve is the exact mean-field calculation, the dash-dot red line is the linear two-body results, and the dash green curve is the quadratic two- plus three-body approximation (Eq.\,\ref{energy}). 
}
\end{figure}
We now consider the non-pertubative case where the interactions are not small as compared to the Rabi coupling strength $\Omega$ corresponding to $\alpha, \gamma \gtrsim1$. In this regime, the energy is evaluated by numerically solving Eq.\,\ref{Eq:spin}, and inserting the result in Eq.\,\ref{eqEMF}. In Fig.\,1, we plot the mean field interaction energy per particle $E_{\rm{int}}/N=(E_{\rm{MF}}(n)/N-\epsilon_-)$ as a function of the parameter $\gamma$ for a detuning $\delta/\Omega=1$ and for scattering lengths that are relevant to our experiment with $^{39}$K. We observe that the interaction energy first grows linearly with the density because of a positive coupling constant in the dressed state. Then it bends down according to the attractive three-body interaction term $g_3$. When $\gamma$ increases above 1, the energy per particle tends to saturate. A Taylor expansion as a function of density is no longer valid for large $\gamma$. In fact, in the limit $\gamma \gg 1$, the system minimizes its energy by choosing the condensate spin state corresponding to a minimal interaction $\cos \theta=\alpha/2\gamma=\Delta g/4\bar{g}$ and a pure two-body interaction is recovered with the coupling constant $g_{\rm{min}}$. In the case $g_{\1\2}=-\sqrt{g_{\1\1}g_{\2\2}}$, $g_{\rm{min}}=0$ and the condensate in the dressed state has an interaction energy that saturates to a constant value at large $\gamma$. 

\begin{figure}
\includegraphics[width=\columnwidth]{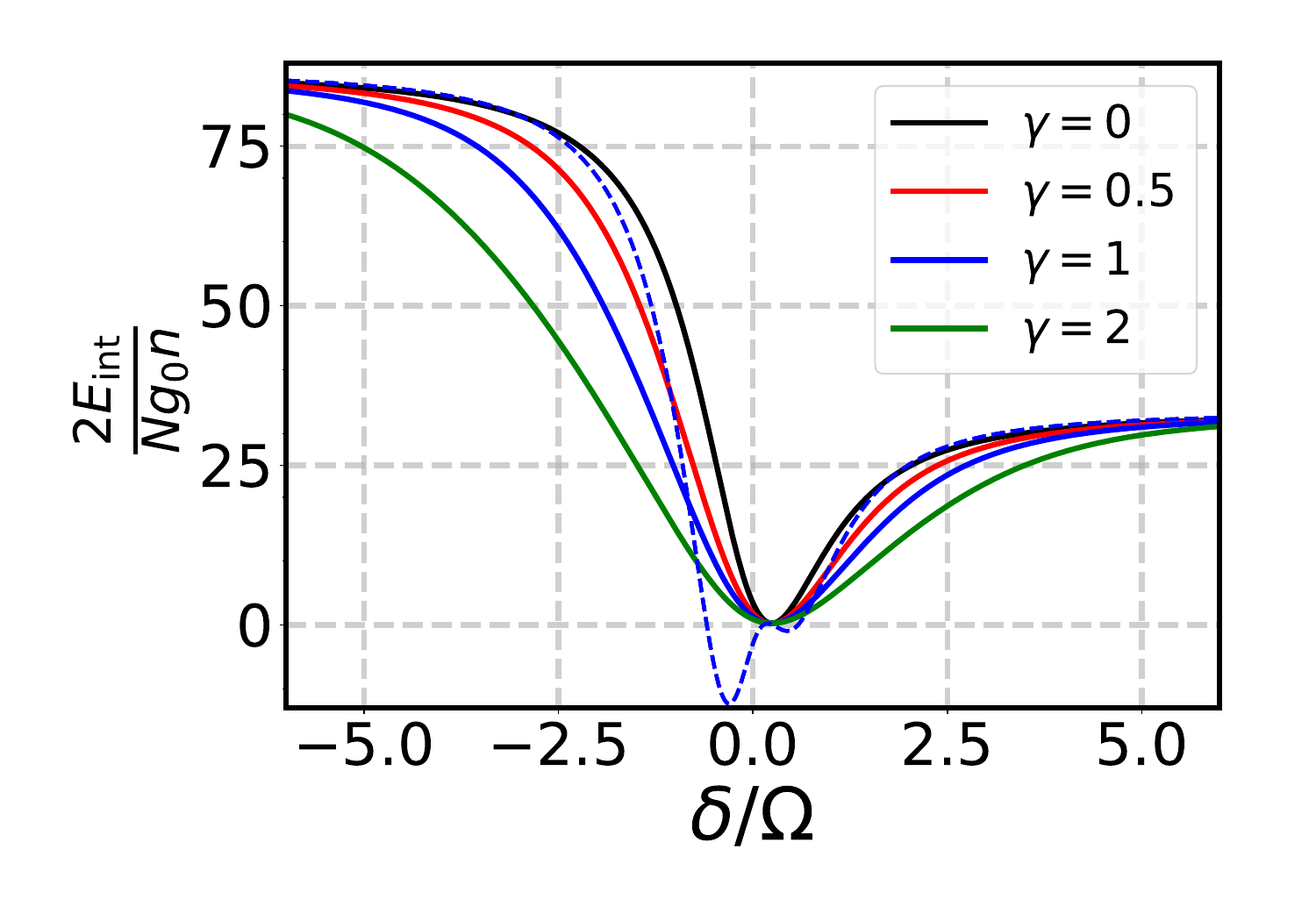}
\caption{\label{fig:energy_delta} 
Interaction energy per particle as a function of the detuning $\delta/\Omega$ for different values of $\gamma$. The scattering lengths are the same as in Fig.\,1. The interaction energy is normalized such that it gives the scattering length in unit of $a_0$ in the regime where the interactions are two-body ($g_0=4\pi \hbar^2 a_0/m$). The bare scattering lengths $a_{\1\1}$ and $a_{\2\2}$ are thus found for large values of $|\delta/\Omega|$. The solid lines correspond to the exact mean-field calculation for different values of $\gamma$. The blue dashed-line corresponds to the two- plus three-body approximation given by Eq.\,\ref{energy} for $\gamma=1$.
}
\end{figure}
The previous results can be experimentally investigated by modifying $\Omega$ in order to change $\gamma$ over a wide range. In Fig. 2, we thus plot the interaction energy as a function of $\delta/\Omega$ for different values of $\gamma$ assuming a homogeneous density $n$. For small values of $\gamma$, the energy is only weakly modified by nonlinear effects, whereas for $\gamma \gtrsim 1$, the effect is large.  The saturation of the interaction energy at large $\gamma$ manifests itself as a broadening of the drop in interaction energy as a function of $\delta/\Omega$. In particular, for $\gamma \gg 1$, the width of the dip increases linearly with $\gamma$. Note that in this regime, the Taylor expansion up to three-body interaction is very wrong. Already for $\gamma=1$, this approximation (blue dashed curve in Fig.\,\ref{fig:energy_delta}) leads to negative mean-field energies, although the interaction energy is always positive for the chosen scattering lengths. 

\section{Experimental results and adjustment}

The experiment starts with a $^{39}$K Bose-Einstein condensate with $N=1.2\times 10^5$ atoms in the $\ket{F=1, m_F=-1}=\ket{\dow}$ spin state in a crossed optical dipole trap with oscillation frequencies $169\times 169 \times 26\,$Hz. The coupling to the state $\ket{F=1, m_F=0}=\ket{\up}$ is performed through a radio frequency (RF) field. The magnetic field $B=56.85\,$G is chosen so that the scattering lengths are $a_{\up\up}=84.3\,$, $a_{\dow\dow}=33.3\,$, $a_{\up\dow}=-53.2\,$, where $a_0$ is the atomic Bohr radius \cite{Tiemann2020}. In short, the measurement idea consists in preparing the dressed state $\ket{-}$ at a specific $\delta/\Omega$ and then monitoring the longitudinal expansion of the cloud once the weak longitudinal confinement is removed. At long times of flight $t$, the expansion is ballistic and we can simply deduce the expansion energy from the relation $E_{\rm{exp}}/N=m\sigma^2/2t^2$, where $\sigma$ is the measured longitudinal rms size of the cloud. For our condition, this expansion energy predominantly reflects the initial interaction energy as the initial kinetic energy is negligible (see below).

The experiment is repeated with different radio frequency powers or equivalently different Rabi coupling frequencies $\Omega/2\pi$ between 0.95 and 30\,kHz. These values of $\Omega$ are chosen in order to explore both the regimes of weak ($\gamma\ll 1$) and strong ($\gamma \gtrsim 1$) nonlinear interaction. Measurements at 0.95\,kHz Rabi frequency are made possible by precise control of detuning $\delta$ down to $\sim$ 100\,Hz. This level of control requires a ppm magnetic field stability and is obtained thanks to a feedforward compensation of both the current in the bias coils and the external magnetic field \cite{Tiengo2025mag}. 

The dressed state $\ket{-}$ is prepared with an adiabatic passage from the state $\ket{\2}$. 
The RF detuning is linearly swept from $9\Omega$ to the desired final value $\delta$. The total sweep time  $\tau$=9\,ms is chosen in order to be adiabatic with respect to the spin degree of freedom even at the lowest $\Omega$ values. In addition, this time is sufficiently large as compared to $1/\omega_\perp$ such that the radial wave-function can adiabatically adapt itself to the interaction change during the sweep \footnote{This is not strictly the case in particular for negative $\delta$ where residual radial oscillations have been observed after the sweep. However, we estimate that these radial oscillations do not significantly affect our longitudinal expansion results.}. 

In the longitudinal direction, the cloud dynamics is much slower, and the interaction change induced by the RF sweep can almost be considered as a sudden quench. However, this approximation is not strictly true. If the RF sweep is done with (resp. without) the longitudinal trap on, the atomic cloud starts to slightly shrink (resp. expands). We reduce this residual longitudinal dynamics by adjusting the time at which the longitudinal trap is turned off during the RF sweep. In principle, the most appropriate time depends on the final value of $\delta/\Omega$ and also on $\Omega$ (through the nonlinear effects). However, we can consider that the additional kinetic energy becomes relevant only when the interaction energy is small, that is, around $\delta/\Omega \approx 0.24$, where $a=a_\textrm{min}\sim 0$. To minimize the longitudinal dynamics during the RF sweep, we thus adjust the longitudinal trap extinction time for each value of $\delta$, ensuring that the induced kinetic energy is minimized when $\delta/\Omega \approx 0.24$. In practice, this is done by searching for the minimal longitudinal expansion at a long time of flight of 300\,ms. We find that the proper trap extinction time varies from 0.5\,ms before the end of the sweep for large $\Omega$ to 5\,ms before for $\Omega = 0.95\,$kHz. The precision of this adjustment allows us to reduce the initial kinetic energy per particle to below 6\,Hz, a negligible quantity compared to the measured interaction energies.

\begin{figure}
\includegraphics[width=\columnwidth]{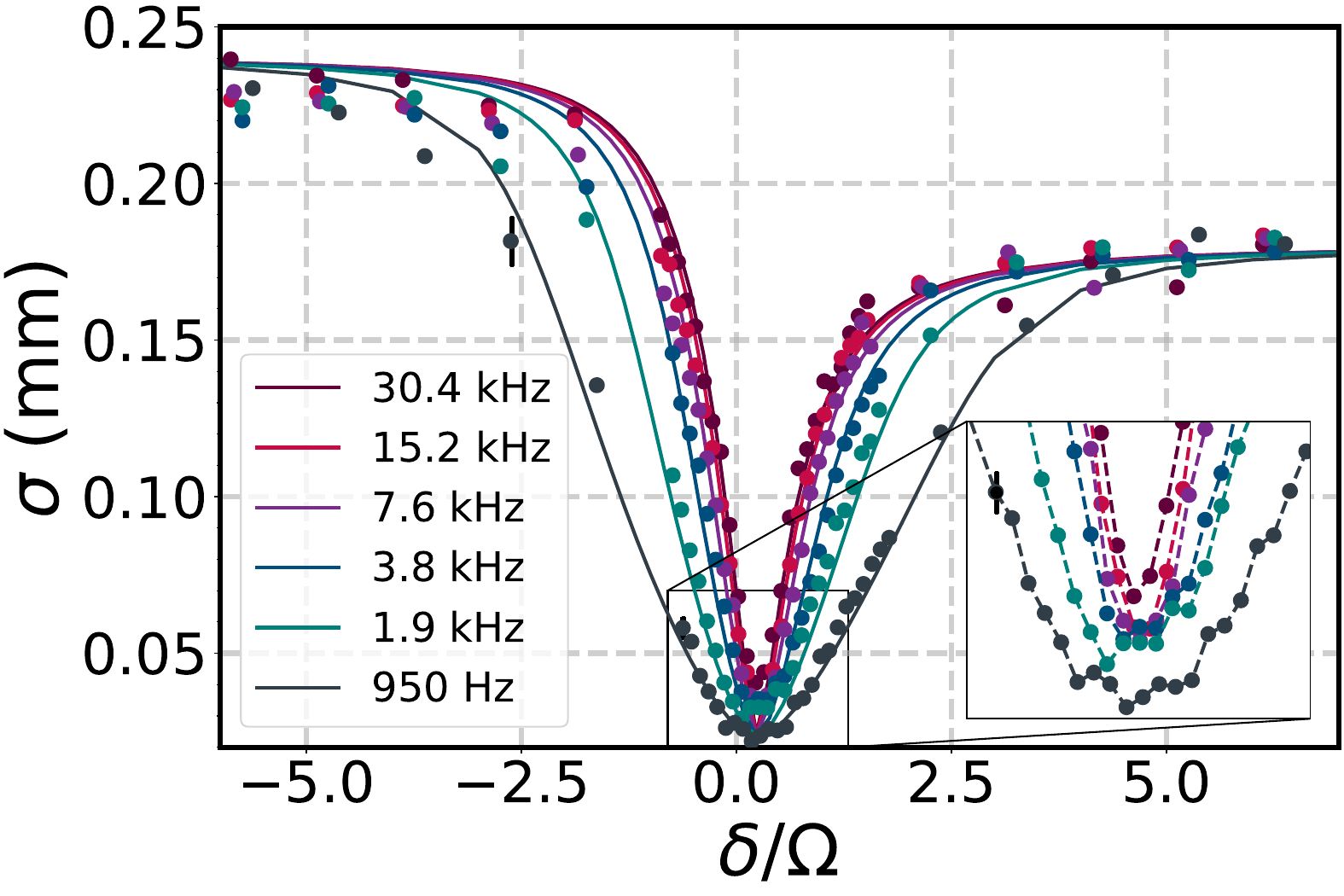}
\caption{\label{fig:results} 
Size after 1D expansion of condensates in the $\ket{-}$ dressed state as a function of the detuning $\delta/\Omega$ and for different values of $\Omega$ (see legend). The points correspond to the experimental measurements and some typical error bars are shown. In the main graph, the solid lines correspond to the mean-field numerical simulations (see text). At $\delta/\Omega\approx 0.24$, the two-body scattering length is already minimal and all theoretical mean-field curves are superimposed. The inset shows an enlargement of the experimental data at low $\delta//\Omega$. The points are linked by dashed lines for clarity. The theoretical lines are omitted in the inset because the theoretical model does not capture the vertical offsets between the curves (see text). 
}
\end{figure}
In Fig.\,3, we plot the longitudinal rms size of the condensate (obtained from a Gaussian fit) after a time of fight $t=62.6\,$ms (during which the radial confinement is kept on) as a function of $\delta/\Omega$ and for different values of $\Omega$. 
Each data set shows a minimum around the value $\delta/\Omega\approx 0.24$ where the two-body coupling constant $g_2$ in the $\ket{-}$ dressed state is minimum. For large and positive $\delta/\Omega$, the expansion corresponds to the one of the condensate in $\ket{\dow}$ with the scattering length $a_{\dow \dow}$. For large and negative $\delta/\Omega$, the expansion is even faster, as the scattering length $a_{\up \up}$ is larger. A striking feature is that the width of the expansion drop is significantly larger as $\Omega$ decreases. This behavior qualitatively follows the theoretical expectation from Fig.\,2 although the initial density is not homogeneous. Even without quantitative analysis, we thus see that in our experiment the strongly nonlinear regime corresponding to $\gamma \gtrsim 1$ is reached. This experimental finding is in line with the estimation $\gamma = 1.6$ for the initial condensate peak density and $\Omega=0.95$\,kHz.


Let us turn to a more quantitative analysis with a numerical simulation of the experimental situation. We have to take into account that the nature of the confinement changes depending on the parameters. Around the expansion minimum, the interaction energy is small and the condensate is in a quasi-1D regime (the radial wavefunction is close to the trap radial ground state). In contrast, for large $\abs{\delta/\Omega}$ and for the initial condensate, the interactions are stronger and the condensate is effectively 3D. In order to account for these two behaviors as well as the dimensional crossover regime, we numerically calculate the radial profile of the spinor condensate wavefunction when the interactions are modified during the sweep. This is done in spinor Gross-Pitaevskii simulations. In this framework, the previously discussed mean-field effects giving rise to the saturating interactions in the equation of states are included. 

In more detail, we assume an adiabatic following of the radial wavefunction,  and thus we search for the radial ground state as a function of the local longitudinal 1D densities. This is done through 2D imaginary-time Gross-Pitaevskii simulations of the spinor wavefunction using the GPELab toolbox \cite{Antoine2014}.  We then calculate the total energy, which includes potential, kinetic, Rabi coupling, and interaction contributions. Finally, we integrate this energy along the longitudinal trap direction, assuming that the density profile is the one of the initial condensate, i.e. a 3D Thomas-Fermi profile. The total energy $E_{\rm{tot}}$ transforms into kinetic energy during expansion. After a significant expansion, the expected size is thus found from the following relation $m\sigma^2/2t^2= E_{\rm{tot}}/N-\hbar \omega_\perp/2-\epsilon_0$, where the last two terms come from the fact that the condensate adiabatically evolves to the noninteracting spinor ground state (with energy $\epsilon_0$) and to the radial ground state during expansion (with energy $\hbar \omega_\perp/2$). 

The results of the above simulations match quite well the experimental findings with no fit parameter (Fig.\,\ref{fig:results}). The width of the minimum expansion drop as a function of $\Omega$ is nicely reproduced, showing that the saturation of the mean-field interaction energy is captured in our mean-field framework. The slightly smaller size at large and negative $\delta/\Omega$ can be attributed to the $\sim 20\%$ losses that are experimentally observed and are due to three-body recombination processes. A smaller atom number reduces the interaction energy as compared to the simulation where no losses are included. 

Looking carefully at the experimental results, one can notice that the observed minimum sizes (at $\delta/\Omega\approx 0.24$) are higher when $\Omega$ increases (see inset in Fig.\,\ref{fig:results}). This behavior is not reproduced in the above mean-field approach, where, at this point, the interaction corresponds to a pure two-body interaction with a coupling constant $g_{\rm{min}}$, which is independent of $\Omega$. We attribute this behavior to beyond mean field effects due to virtual excitations to the $\ket{+}$ states, leading to a slight increase of the minimal interaction energy when $\Omega$ increases \cite{Lavoine2021, Tiengo2025}. In the context of this paper, we do not focus on this residual energy shift, as its magnitude is much lower than the dominant mean-field effects studied above. Its quantitative measurement is demanding and would require an improved experimental control on the initial kinetic energy of the cloud.

\section{Conclusion}
In this paper, we have shown that the density dependence of the equation of state of a coherently driven two-component condensate can be tuned. For some parameters, the interaction term evolves from a two-body repulsive interaction at low densities, through a regime that can be described with additional three-body interactions at intermediate densities, to yet another regime of two-body interaction with a reduced two-body constant at large densities. This second two-body interaction can be positive or negative, depending on the initial spin-dependent scattering lengths. If it is close to zero (as in our work), then it leads to an unusual equation of state with an energy saturation as a function of the density. 

Our findings open the way to the study of novel nonlinear effects in the physics of condensates. Already, at equilibrium, we anticipate density profiles that differ from the usual Thomas-Fermi profiles. The condensate excitation dynamics (vortices, solitons, low energy excitation modes) will also be modified. For example, in our configuration, the speed of sound is expected to decrease at high densities, in contrast to the usual behavior for a Bose-Einstein condensate with repulsive interactions.

\begin{acknowledgments}
 This  research  was  supported  by  CNRS,  Minist\`ere  de  l'Enseignement  Sup\'erieur  et  de  la  Recherche,  Quantum Paris-Saclay, R\'egion Ile-de-France  in  the  framework  of  Domaine d'Int\'er\^et Majeur Quantip (H3B), France 2030 in the framework of PEPR Quantique (ANR-23-PETQ-0001), the Simons Foundation (Award No. 563916,  localization of waves), France and Chicago collaborating in the sciences.
\end{acknowledgments}

\appendix


\bibliography{references_sat}

\begin{thebibliography}{19}%
\makeatletter
\providecommand \@ifxundefined [1]{%
 \@ifx{#1\undefined}
}%
\providecommand \@ifnum [1]{%
 \ifnum #1\expandafter \@firstoftwo
 \else \expandafter \@secondoftwo
 \fi
}%
\providecommand \@ifx [1]{%
 \ifx #1\expandafter \@firstoftwo
 \else \expandafter \@secondoftwo
 \fi
}%
\providecommand \natexlab [1]{#1}%
\providecommand \enquote  [1]{``#1''}%
\providecommand \bibnamefont  [1]{#1}%
\providecommand \bibfnamefont [1]{#1}%
\providecommand \citenamefont [1]{#1}%
\providecommand \href@noop [0]{\@secondoftwo}%
\providecommand \href [0]{\begingroup \@sanitize@url \@href}%
\providecommand \@href[1]{\@@startlink{#1}\@@href}%
\providecommand \@@href[1]{\endgroup#1\@@endlink}%
\providecommand \@sanitize@url [0]{\catcode `\\12\catcode `\$12\catcode
  `\&12\catcode `\#12\catcode `\^12\catcode `\_12\catcode `\%12\relax}%
\providecommand \@@startlink[1]{}%
\providecommand \@@endlink[0]{}%
\providecommand \url  [0]{\begingroup\@sanitize@url \@url }%
\providecommand \@url [1]{\endgroup\@href {#1}{\urlprefix }}%
\providecommand \urlprefix  [0]{URL }%
\providecommand \Eprint [0]{\href }%
\providecommand \doibase [0]{https://doi.org/}%
\providecommand \selectlanguage [0]{\@gobble}%
\providecommand \bibinfo  [0]{\@secondoftwo}%
\providecommand \bibfield  [0]{\@secondoftwo}%
\providecommand \translation [1]{[#1]}%
\providecommand \BibitemOpen [0]{}%
\providecommand \bibitemStop [0]{}%
\providecommand \bibitemNoStop [0]{.\EOS\space}%
\providecommand \EOS [0]{\spacefactor3000\relax}%
\providecommand \BibitemShut  [1]{\csname bibitem#1\endcsname}%
\let\auto@bib@innerbib\@empty
\bibitem [{\citenamefont {Strogatz}(2024)}]{Strogatz2024}%
  \BibitemOpen
  \bibfield  {author} {\bibinfo {author} {\bibfnamefont {S.~H.}\ \bibnamefont
  {Strogatz}},\ }\href@noop {} {\emph {\bibinfo {title} {Nonlinear Dynamics and
  Chaos: With Applications to Physics, Biology, Chemistry, and Engineering}}},\
  \bibinfo {edition} {3rd}\ ed.\ (\bibinfo  {publisher} {CRC Press},\ \bibinfo
  {address} {Boca Raton, FL},\ \bibinfo {year} {2024})\ p.\ \bibinfo {pages}
  {616}\BibitemShut {NoStop}%
\bibitem [{\citenamefont {Shen}(1984)}]{Shen1984}%
  \BibitemOpen
  \bibfield  {author} {\bibinfo {author} {\bibfnamefont {Y.-R.}\ \bibnamefont
  {Shen}},\ }\href@noop {} {\emph {\bibinfo {title} {The Principles of
  Nonlinear Optics}}}\ (\bibinfo  {publisher} {Wiley-Interscience},\ \bibinfo
  {address} {New York},\ \bibinfo {year} {1984})\BibitemShut {NoStop}%
\bibitem [{\citenamefont {Boyd}(2020)}]{Boyd2020}%
  \BibitemOpen
  \bibfield  {author} {\bibinfo {author} {\bibfnamefont {R.~W.}\ \bibnamefont
  {Boyd}},\ }\href@noop {} {\emph {\bibinfo {title} {Nonlinear Optics}}},\
  \bibinfo {edition} {4th}\ ed.\ (\bibinfo  {publisher} {Academic Press},\
  \bibinfo {year} {2020})\BibitemShut {NoStop}%
\bibitem [{\citenamefont {Pitaevskii}\ and\ \citenamefont
  {Stringari}(2003)}]{Pitaevskii2003}%
  \BibitemOpen
  \bibfield  {author} {\bibinfo {author} {\bibfnamefont {P.}~\bibnamefont
  {Pitaevskii}, \bibfnamefont {Lev}}\ and\ \bibinfo {author} {\bibfnamefont
  {S.}~\bibnamefont {Stringari}},\ }\href@noop {} {\emph {\bibinfo {title}
  {{B}ose–{E}instein Condensation}}}\ (\bibinfo  {publisher} {Clarendon
  Press, Oxford Univ. Press},\ \bibinfo {address} {Oxford, UK},\ \bibinfo
  {year} {2003})\ p.\ \bibinfo {pages} {382}\BibitemShut {NoStop}%
\bibitem [{\citenamefont {Lee}\ \emph {et~al.}(1957)\citenamefont {Lee},
  \citenamefont {Huang},\ and\ \citenamefont {Yang}}]{Lee1957}%
  \BibitemOpen
  \bibfield  {author} {\bibinfo {author} {\bibfnamefont {T.~D.}\ \bibnamefont
  {Lee}}, \bibinfo {author} {\bibfnamefont {K.}~\bibnamefont {Huang}},\ and\
  \bibinfo {author} {\bibfnamefont {C.~N.}\ \bibnamefont {Yang}},\ }\bibfield
  {title} {\bibinfo {title} {Eigenvalues and eigenfunctions of a bose system of
  hard spheres and its low-temperature properties},\ }\href
  {https://doi.org/10.1103/PhysRev.106.1135} {\bibfield  {journal} {\bibinfo
  {journal} {Physical Review}\ }\textbf {\bibinfo {volume} {106}},\ \bibinfo
  {pages} {1135} (\bibinfo {year} {1957})}\BibitemShut {NoStop}%
\bibitem [{\citenamefont {Petrov}(2015)}]{petrov2015}%
  \BibitemOpen
  \bibfield  {author} {\bibinfo {author} {\bibfnamefont {D.~S.}\ \bibnamefont
  {Petrov}},\ }\bibfield  {title} {\bibinfo {title} {Quantum mechanical
  stabilization of a collapsing {B}ose–{B}ose mixture},\ }\href
  {https://doi.org/10.1103/PhysRevLett.115.155302} {\bibfield  {journal}
  {\bibinfo  {journal} {Physical Review Letters}\ }\textbf {\bibinfo {volume}
  {115}},\ \bibinfo {pages} {155302} (\bibinfo {year} {2015})}\BibitemShut
  {NoStop}%
\bibitem [{\citenamefont {Cabrera}\ \emph {et~al.}(2018)\citenamefont
  {Cabrera}, \citenamefont {Tanzi}, \citenamefont {Sanz}, \citenamefont
  {Naylor}, \citenamefont {Thomas}, \citenamefont {Cheiney},\ and\
  \citenamefont {Tarruell}}]{cabrera2018}%
  \BibitemOpen
  \bibfield  {author} {\bibinfo {author} {\bibfnamefont {C.~R.}\ \bibnamefont
  {Cabrera}}, \bibinfo {author} {\bibfnamefont {L.}~\bibnamefont {Tanzi}},
  \bibinfo {author} {\bibfnamefont {J.}~\bibnamefont {Sanz}}, \bibinfo {author}
  {\bibfnamefont {B.}~\bibnamefont {Naylor}}, \bibinfo {author} {\bibfnamefont
  {P.}~\bibnamefont {Thomas}}, \bibinfo {author} {\bibfnamefont
  {P.}~\bibnamefont {Cheiney}},\ and\ \bibinfo {author} {\bibfnamefont
  {L.}~\bibnamefont {Tarruell}},\ }\bibfield  {title} {\bibinfo {title}
  {Quantum liquid droplets in a mixture of {B}ose–{E}instein condensates},\
  }\href {https://doi.org/10.1126/science.aao5686} {\bibfield  {journal}
  {\bibinfo  {journal} {Science}\ }\textbf {\bibinfo {volume} {359}},\ \bibinfo
  {pages} {301} (\bibinfo {year} {2018})}\BibitemShut {NoStop}%
\bibitem [{\citenamefont {Braaten}\ and\ \citenamefont
  {Nieto}(1997)}]{Braaten1997}%
  \BibitemOpen
  \bibfield  {author} {\bibinfo {author} {\bibfnamefont {E.}~\bibnamefont
  {Braaten}}\ and\ \bibinfo {author} {\bibfnamefont {A.}~\bibnamefont
  {Nieto}},\ }\bibfield  {title} {\bibinfo {title} {Renormalization effects in
  a dilute {B}ose gas},\ }\href {https://doi.org/10.1103/PhysRevB.55.8090}
  {\bibfield  {journal} {\bibinfo  {journal} {Phys. Rev. B}\ }\textbf {\bibinfo
  {volume} {55}},\ \bibinfo {pages} {8090} (\bibinfo {year}
  {1997})}\BibitemShut {NoStop}%
\bibitem [{\citenamefont {K\"ohler}(2002)}]{Kohler02}%
  \BibitemOpen
  \bibfield  {author} {\bibinfo {author} {\bibfnamefont {T.}~\bibnamefont
  {K\"ohler}},\ }\bibfield  {title} {\bibinfo {title} {Three-body problem in a
  dilute {B}ose-{E}instein condensate},\ }\href
  {https://doi.org/10.1103/PhysRevLett.89.210404} {\bibfield  {journal}
  {\bibinfo  {journal} {Phys. Rev. Lett.}\ }\textbf {\bibinfo {volume} {89}},\
  \bibinfo {pages} {210404} (\bibinfo {year} {2002})}\BibitemShut {NoStop}%
\bibitem [{\citenamefont {Mestrom}\ \emph {et~al.}(2019)\citenamefont
  {Mestrom}, \citenamefont {Colussi}, \citenamefont {Secker},\ and\
  \citenamefont {Kokkelmans}}]{Mestrom19}%
  \BibitemOpen
  \bibfield  {author} {\bibinfo {author} {\bibfnamefont {P.~M.~A.}\
  \bibnamefont {Mestrom}}, \bibinfo {author} {\bibfnamefont {V.~E.}\
  \bibnamefont {Colussi}}, \bibinfo {author} {\bibfnamefont {T.}~\bibnamefont
  {Secker}},\ and\ \bibinfo {author} {\bibfnamefont {S.~J. J. M.~F.}\
  \bibnamefont {Kokkelmans}},\ }\bibfield  {title} {\bibinfo {title}
  {Scattering hypervolume for ultracold bosons from weak to strong
  interactions},\ }\href {https://doi.org/10.1103/PhysRevA.100.050702}
  {\bibfield  {journal} {\bibinfo  {journal} {Phys. Rev. A}\ }\textbf {\bibinfo
  {volume} {100}},\ \bibinfo {pages} {050702} (\bibinfo {year}
  {2019})}\BibitemShut {NoStop}%
\bibitem [{\citenamefont {Sanz}\ \emph {et~al.}(2022)\citenamefont {Sanz},
  \citenamefont {Fr\"olian}, \citenamefont {Chisholm}, \citenamefont
  {Cabrera},\ and\ \citenamefont {Tarruell}}]{sanz2022}%
  \BibitemOpen
  \bibfield  {author} {\bibinfo {author} {\bibfnamefont {J.}~\bibnamefont
  {Sanz}}, \bibinfo {author} {\bibfnamefont {A.}~\bibnamefont {Fr\"olian}},
  \bibinfo {author} {\bibfnamefont {C.~S.}\ \bibnamefont {Chisholm}}, \bibinfo
  {author} {\bibfnamefont {C.~R.}\ \bibnamefont {Cabrera}},\ and\ \bibinfo
  {author} {\bibfnamefont {L.}~\bibnamefont {Tarruell}},\ }\bibfield  {title}
  {\bibinfo {title} {Interaction control and bright solitons in
  coherently-coupled {B}ose-{E}instein condensates},\ }\href
  {https://doi.org/10.1103/PhysRevLett.128.013201} {\bibfield  {journal}
  {\bibinfo  {journal} {Physical Review Letters}\ }\textbf {\bibinfo {volume}
  {128}},\ \bibinfo {pages} {013201} (\bibinfo {year} {2022})}\BibitemShut
  {NoStop}%
\bibitem [{\citenamefont {Tiengo}\ \emph
  {et~al.}(2025{\natexlab{a}})\citenamefont {Tiengo}, \citenamefont {Eid},\
  and\ \citenamefont {Bourdel}}]{Tiengo2025}%
  \BibitemOpen
  \bibfield  {author} {\bibinfo {author} {\bibfnamefont {S.}~\bibnamefont
  {Tiengo}}, \bibinfo {author} {\bibfnamefont {R.}~\bibnamefont {Eid}},\ and\
  \bibinfo {author} {\bibfnamefont {T.}~\bibnamefont {Bourdel}},\ }\bibfield
  {title} {\bibinfo {title} {Three-body interactions in {R}abi-coupled bose
  gases: A perturbative approach},\ }\href
  {https://doi.org/10.1103/PhysRevA.111.053319} {\bibfield  {journal} {\bibinfo
   {journal} {Phys. Rev. A}\ }\textbf {\bibinfo {volume} {111}},\ \bibinfo
  {pages} {053319} (\bibinfo {year} {2025}{\natexlab{a}})}\BibitemShut
  {NoStop}%
\bibitem [{\citenamefont {Hammond}\ \emph {et~al.}(2022)\citenamefont
  {Hammond}, \citenamefont {Lavoine},\ and\ \citenamefont
  {Bourdel}}]{Hammond2022}%
  \BibitemOpen
  \bibfield  {author} {\bibinfo {author} {\bibfnamefont {A.}~\bibnamefont
  {Hammond}}, \bibinfo {author} {\bibfnamefont {L.}~\bibnamefont {Lavoine}},\
  and\ \bibinfo {author} {\bibfnamefont {T.}~\bibnamefont {Bourdel}},\
  }\bibfield  {title} {\bibinfo {title} {Tunable three-body interactions in
  driven two-component {B}ose-{E}instein condensates},\ }\href
  {https://doi.org/10.1103/PhysRevLett.128.083401} {\bibfield  {journal}
  {\bibinfo  {journal} {Phys. Rev. Lett.}\ }\textbf {\bibinfo {volume} {128}},\
  \bibinfo {pages} {083401} (\bibinfo {year} {2022})}\BibitemShut {NoStop}%
\bibitem [{\citenamefont {Farolfi}\ \emph {et~al.}(2021)\citenamefont
  {Farolfi}, \citenamefont {Zenesini}, \citenamefont {Cominotti}, \citenamefont
  {Trypogeorgos}, \citenamefont {Recati}, \citenamefont {Lamporesi},\ and\
  \citenamefont {Ferrari}}]{Farolfi2021}%
  \BibitemOpen
  \bibfield  {author} {\bibinfo {author} {\bibfnamefont {A.}~\bibnamefont
  {Farolfi}}, \bibinfo {author} {\bibfnamefont {A.}~\bibnamefont {Zenesini}},
  \bibinfo {author} {\bibfnamefont {R.}~\bibnamefont {Cominotti}}, \bibinfo
  {author} {\bibfnamefont {D.}~\bibnamefont {Trypogeorgos}}, \bibinfo {author}
  {\bibfnamefont {A.}~\bibnamefont {Recati}}, \bibinfo {author} {\bibfnamefont
  {G.}~\bibnamefont {Lamporesi}},\ and\ \bibinfo {author} {\bibfnamefont
  {G.}~\bibnamefont {Ferrari}},\ }\bibfield  {title} {\bibinfo {title}
  {Manipulation of an elongated internal {J}osephson junction of bosonic
  atoms},\ }\href {https://doi.org/10.1103/PhysRevA.104.023326} {\bibfield
  {journal} {\bibinfo  {journal} {Phys. Rev. A}\ }\textbf {\bibinfo {volume}
  {104}},\ \bibinfo {pages} {023326} (\bibinfo {year} {2021})}\BibitemShut
  {NoStop}%
\bibitem [{\citenamefont {Tiemann}\ \emph {et~al.}(2020)\citenamefont
  {Tiemann}, \citenamefont {Gersema}, \citenamefont {Voges}, \citenamefont
  {Hartmann}, \citenamefont {Zenesini},\ and\ \citenamefont
  {Ospelkaus}}]{Tiemann2020}%
  \BibitemOpen
  \bibfield  {author} {\bibinfo {author} {\bibfnamefont {E.}~\bibnamefont
  {Tiemann}}, \bibinfo {author} {\bibfnamefont {P.}~\bibnamefont {Gersema}},
  \bibinfo {author} {\bibfnamefont {K.~K.}\ \bibnamefont {Voges}}, \bibinfo
  {author} {\bibfnamefont {T.}~\bibnamefont {Hartmann}}, \bibinfo {author}
  {\bibfnamefont {A.}~\bibnamefont {Zenesini}},\ and\ \bibinfo {author}
  {\bibfnamefont {S.}~\bibnamefont {Ospelkaus}},\ }\bibfield  {title} {\bibinfo
  {title} {Beyond {B}orn-{O}ppenheimer approximation in ultracold atomic
  collisions},\ }\href {https://doi.org/10.1103/PhysRevResearch.2.013366}
  {\bibfield  {journal} {\bibinfo  {journal} {Phys. Rev. Res.}\ }\textbf
  {\bibinfo {volume} {2}},\ \bibinfo {pages} {013366} (\bibinfo {year}
  {2020})}\BibitemShut {NoStop}%
\bibitem [{\citenamefont {Tiengo}\ \emph
  {et~al.}(2025{\natexlab{b}})\citenamefont {Tiengo}, \citenamefont {Eid},
  \citenamefont {Apfel}, \citenamefont {Brulin},\ and\ \citenamefont
  {Bourdel}}]{Tiengo2025mag}%
  \BibitemOpen
  \bibfield  {author} {\bibinfo {author} {\bibfnamefont {S.}~\bibnamefont
  {Tiengo}}, \bibinfo {author} {\bibfnamefont {R.}~\bibnamefont {Eid}},
  \bibinfo {author} {\bibfnamefont {M.}~\bibnamefont {Apfel}}, \bibinfo
  {author} {\bibfnamefont {G.}~\bibnamefont {Brulin}},\ and\ \bibinfo {author}
  {\bibfnamefont {T.}~\bibnamefont {Bourdel}},\ }\bibfield  {title} {\bibinfo
  {title} {A simple magnetic field stabilization technique for atomic
  {B}ose-{E}instein condensate experiments},\ }\href
  {https://doi.org/10.1063/5.0258855} {\bibfield  {journal} {\bibinfo
  {journal} {Review of Scientific Instruments}\ }\textbf {\bibinfo {volume}
  {96}},\ \bibinfo {pages} {063201} (\bibinfo {year}
  {2025}{\natexlab{b}})}\BibitemShut {NoStop}%
\bibitem [{Note1()}]{Note1}%
  \BibitemOpen
  \bibinfo {note} {This is not strictly the case in particular for negative
  $\delta $ where residual radial oscillations have been observed after the
  sweep. However, we estimate that these radial oscillations do not
  significantly affect our longitudinal expansion results.}\BibitemShut {Stop}%
\bibitem [{\citenamefont {Antoine}\ and\ \citenamefont
  {Duboscq}(2014)}]{Antoine2014}%
  \BibitemOpen
  \bibfield  {author} {\bibinfo {author} {\bibfnamefont {X.}~\bibnamefont
  {Antoine}}\ and\ \bibinfo {author} {\bibfnamefont {R.}~\bibnamefont
  {Duboscq}},\ }\bibfield  {title} {\bibinfo {title} {Gpelab, a {Matlab}
  toolbox to solve {G}ross–{0}pitaevskii equations {I}: computation of
  stationary solutions},\ }\href@noop {} {\bibfield  {journal} {\bibinfo
  {journal} {Computer Physics Communications}\ }\textbf {\bibinfo {volume}
  {185}},\ \bibinfo {pages} {2969} (\bibinfo {year} {2014})}\BibitemShut
  {NoStop}%
\bibitem [{\citenamefont {Lavoine}\ \emph {et~al.}(2021)\citenamefont
  {Lavoine}, \citenamefont {Hammond}, \citenamefont {Recati}, \citenamefont
  {Petrov},\ and\ \citenamefont {Bourdel}}]{Lavoine2021}%
  \BibitemOpen
  \bibfield  {author} {\bibinfo {author} {\bibfnamefont {L.}~\bibnamefont
  {Lavoine}}, \bibinfo {author} {\bibfnamefont {A.}~\bibnamefont {Hammond}},
  \bibinfo {author} {\bibfnamefont {A.}~\bibnamefont {Recati}}, \bibinfo
  {author} {\bibfnamefont {D.~S.}\ \bibnamefont {Petrov}},\ and\ \bibinfo
  {author} {\bibfnamefont {T.}~\bibnamefont {Bourdel}},\ }\bibfield  {title}
  {\bibinfo {title} {Beyond-mean-field effects in {R}abi-coupled two-component
  {B}ose-{E}instein condensate},\ }\href
  {https://doi.org/10.1103/PhysRevLett.127.203402} {\bibfield  {journal}
  {\bibinfo  {journal} {Phys. Rev. Lett.}\ }\textbf {\bibinfo {volume} {127}},\
  \bibinfo {pages} {203402} (\bibinfo {year} {2021})}\BibitemShut {NoStop}%
\end{thebibliography}%

\end{document}